\documentclass[12pt,preprint,psfig]{aastex}

\usepackage{emulateapj5}

\def\ltsima{$\; \buildrel < \over \sim \;$}

\def\lsim{\lower.5ex\hbox{\ltsima}}
\def\gtsima{$\; \buildrel > \over \sim \;$}
\def\gsim{\lower.5ex\hbox{\gtsima}}

\begin{document}
\title{Constraints on the Process that Regulates the Growth of Supermassive
Black Holes Based on the Intrinsic Scatter in the $M_{\rm bh}$--$\sigma$
Relation}

\author{J. Stuart B. Wyithe\altaffilmark{1}, Abraham Loeb\altaffilmark{2}}

\email{swyithe@physics.unimelb.edu.au; aloeb@cfa.harvard.edu}

\altaffiltext{1}{University of Melbourne, Parkville, Victoria, Australia}

\altaffiltext{2}{Astronomy department, Harvard University, 60 Garden
St., Cambridge, MA 02138, USA}

\begin{abstract}
\noindent 
We show that the observed scatter in the relations between the mass of
supermassive black holes (SMBHs), $M_{\rm bh}$, and the velocity dispersion
$\sigma_{\rm sph}$ or mass $M_{\rm sph}$ of their host spheroid, place
interesting constraints on the process that regulates SMBH growth in
galaxies.  When combined with the observed properties of early-type SDSS
galaxies, the observed intrinsic scatters imply that SMBH growth is
regulated by the spheroid velocity dispersion rather than its mass. The
$M_{\rm bh}$--$M_{\rm sph}$ relation is therefore a by-product of a more
fundamental $M_{\rm bh}$--$\sigma_{\rm sph}$ relation. We construct a
theoretical model for the scatter among baryon modified dark matter halo
profiles, out of which we generate a population of spheroid hosts and show
that these naturally lead to a relation between effective radius and
velocity dispersion of the form $R_{\rm sph}\approx 6{\rm kpc}(\sigma_{\rm
sph}/200\mbox{km}\,\mbox{s}^{-1})^{1.5}$ with a scatter of $\sim$0.2dex, in
agreement with the corresponding projection of the fundamental plane for
early type galaxies in SDSS.  At the redshift of formation, our model
predicts the minimum scatter that SMBHs can have at fixed velocity
dispersion or spheroid mass under different formation scenarios. We also
estimate the additional scatter that is introduced into these relations
through collisionless mergers of purely stellar spheroids at $z<1$.  We
find that the observed scatter in the $M_{\rm bh}$--$\sigma_{\rm sph}$ and
$M_{\rm bh}$--$M_{\rm sph}$ relations preclude the properties of dark
matter halos from being the governing factor in SMBH growth. The apparent
relation between halo and SMBH mass is merely a reflection of the fact that
massive halos tend to host massive stellar spheroids (albeit with a large
scatter owing to the variance in formation histories). Finally, we show
that SMBH growth governed by the properties of the host spheroid can lead
to the observed values of scatter in the $M_{\rm bh}$--$\sigma_{\rm sph}$
and $M_{\rm bh}$--$M_{\rm sph}$ relations, only if the SMBH growth is
limited by momentum or energy feedback {\em over the dynamical time} of the
host spheroid.

\end{abstract}

\keywords{cosmology: theory - galaxies: formation}

\section{Introduction}

Supermassive black holes (SMBHs) are a ubiquitous constituent of spheroids
in nearby galaxies (e.g. Kormandy \& Richstone~1995). A decade ago it
became apparent that the masses of SMBHs correlate with the luminosity of
the host spheroid (Kormandy \& Richstone~1995). Subsequently, other
correlations with substantially less intrinsic scatter have been
discovered. These include correlations between the SMBH mass ($M_{\rm bh}$),
and the mass, $M_{\rm sph}$ (Magorrian et al.~1998; Haering \& Rix~2004),
stellar velocity dispersion, $\sigma_{\rm sph}$ (Ferrarese \&
Merritt~2000; Gebhardt et al.~2000), and concentration (Graham et
al.~2002) of its host spheroid. The tightest relation, with intrinsic
scatter of $\delta\sim0.275\pm0.05$dex, appears to be between SMBH mass and
velocity dispersion (Tremaine et al.~2002; Wyithe~2005).

These relations must hold important clues about the astrophysics that
regulates the growth of a SMBH and its impact on galaxy
formation. While much attention was dedicated towards interpreting the
power-law slope of the $M_{\rm bh}$--$\sigma_{\rm sph}$ relation via a
slew of analytic, semi-analytic and numerical attempts to reproduce it
(e.g. Silk \& Rees~1998; Wyithe \& Loeb~2003; King~2003;
Miralda-Escude \& Kollmeier~2005; Adams et al.~2003; Sazonov et
al.~2005; Di Matteo et al.~2005), little attention has been directed
towards interpreting the constraints that its {\it intrinsic scatter}
might place on our understanding of SMBH growth (Robertson et
al.~2005). Moreover, agreement between data and theory is a necessary
but not sufficient condition. A model that reproduces the observations is not
necessarily unique. The various successful attempts to model the
quasar luminosity function assuming different physical models attest
to this fact. Our goals in this paper are to use the observed scatter
in the $M_{\rm bh}$--$\sigma_{\rm sph}$ and $M_{\rm bh}$--$M_{\rm
sph}$ relations in addition to their power-law slope as a diagnostic
of SMBH formation physics, and to constrain a range of possible models
of SMBH formation.

Throughout the paper we adopt the set of cosmological parameters
determined by the {\em Wilkinson Microwave Anisotropy Probe} (WMAP,
Spergel et al. 2003), namely mass density parameters of
$\Omega_{m}=0.27$ in matter, $\Omega_{b}=0.044$ in baryons,
$\Omega_\Lambda=0.73$ in a cosmological constant, and a Hubble
constant of $H_0=71~{\rm km\,s^{-1}\,Mpc^{-1}}$.

\section{Intrinsic scatter in the observed 
$M_{\rm bh}$--$\sigma_{\rm sph}$ 
and $M_{\rm bh}$--$M_{\rm sph}$ relations}

Tremaine et al.~(2002) have compiled a list of spheroids with reliable
determinations of both SMBH mass and central velocity dispersion (defined
as the luminosity weighted dispersion in a slit aperture of half length
$R_{\rm sph}$). These SMBHs show a tight correlation between $M_{\rm bh}$
and $\sigma_{\rm sph}$ (Gebhardt et al.~2000; Ferrarese \&
Merritt~2000). Recently, Wyithe~(2005) found that the $M_{\rm
bh}$--$\sigma_{\rm sph}$ relation within the sample of Tremaine et
al.~(2002) shows evidence for a deviation from a pure power-law behavior,
and obtained an a-posteriori probability distribution for the scatter in
the relation of $\delta=0.28\pm0.05$. The best-fit log-quadratic $M_{\rm
bh}$--$\sigma_{\rm sph}$ takes the form
\begin{eqnarray}
\label{Msig}
\nonumber
\log_{10}(M_{\rm bh})&=&\alpha+\beta\log_{10}\left(\sigma_{\rm sph}/200\mbox{km/s}\right)\\
 &+&\beta_2\left[\log_{10}\left(\sigma_{\rm sph}/200\mbox{km/s}\right)\right]^2,
\end{eqnarray}
where $\beta=4.2\pm0.37$ corresponds to the traditional power-law relation
and $\beta_{2}=1.6\pm1.3$ quantifies the departure of the local SMBH sample
from a pure power-law.

A correlation has also been found between the mass of the spheroid
component, $M_{\rm sph}$, and the SMBH (Magorrian et al.~1998).  The sample
compiled by Haering \& Rix~(2004) (which overlaps
predominantly with the Tremaine et al.~2002 sample) was studied by
Wyithe~(2005) who found that like the $M_{\rm bh}$--$\sigma_{\rm sph}$
relation, the $M_{\rm bh}$--$M_{\rm sph}$ relation may depart from a
uniform power-law, and that the scatter in the $M_{\rm bh}$--$M_{\rm sph}$
relation is larger than in the $M_{\rm bh}$--$\sigma_{\rm sph}$ relation by
50\% ($\delta_{\rm sph}=0.41\pm0.07$). 
The best-fit log-quadratic $M_{\rm bh}$--$M_{\rm sph}$ takes the form
\begin{eqnarray}
\label{MM}
\nonumber
\log_{10}(M_{\rm bh})&=&\alpha_{\rm sph}+\beta_{\rm sph}\log_{10}\left(M_{\rm sph}/10^{11}M_\odot\right)\\
 &+&\beta_{2,{\rm sph}}\left[\log_{10}\left(M_{\rm sph}/10^{11}M_\odot\right)\right]^2,
\end{eqnarray}
where $\beta_{\rm sph}=1.15\pm0.18$ and $\beta_{2,{\rm sph}}=1.12\pm0.14$.

In this paper we use the values of intrinsic scatter ($\delta$ and
$\delta_{\rm sph}$) in the local SMBH sample to place constraints on the
process that regulates SMBH growth in galaxies.

\section{Which Relation is More Fundamental: 
$M_{\rm bh}$--$\sigma_{\rm sph}$ or $M_{\rm bh}$--$M_{\rm sph}$?}

\begin{figure*}[t]
\epsscale{1.8} \plotone{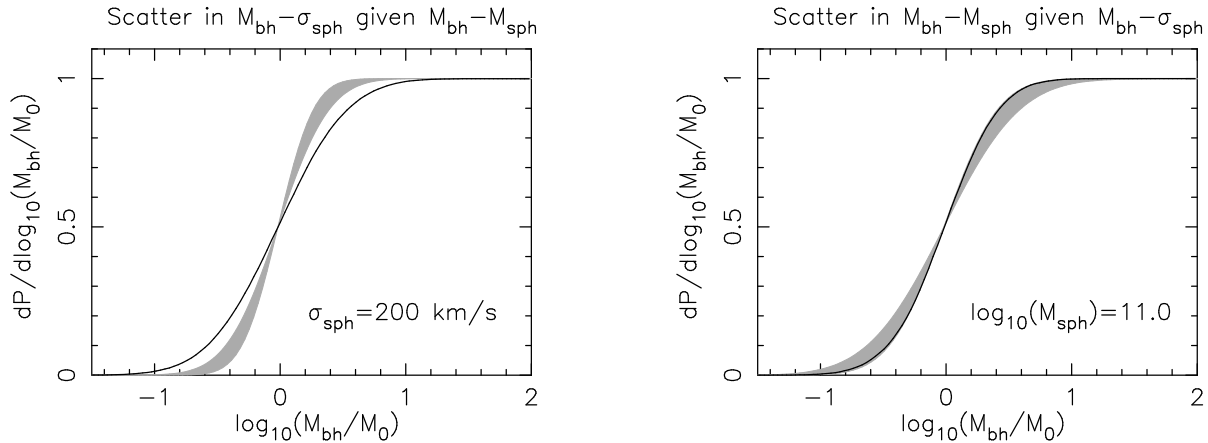}
\caption{\label{fig1} {\em Left:} The predicted intrinsic cumulative
distribution of residuals ($\log_{10}(M_{\rm bh}/M_0)$, where $M_0$ is the
mass corresponding to the mean local relation) in $M_{\rm
bh}$ at constant $\sigma_{\rm sph}=200~{\rm
km~s^{-1}}$. In this case the observed $M_{\rm bh}$--$M_{\rm sph}$ relation with
intrinsic scatter of $\delta_{\rm sph}=0.41\pm0.07$dex was assumed to be
fundamental. {\em Right:} The predicted intrinsic residual scatter in the
local $M_{\rm bh}$ at constant $M_{\rm sph}=10^{11}M_{\odot}$ assuming the
observed $M_{\rm bh}$--$\sigma_{\rm sph}$ relation with intrinsic
scatter of $\delta=0.275\pm0.05$dex to be fundamental. In each panel the
range of the observed scatter (1-sigma) is shown as the shaded grey region
for comparison. Zero scatter would have been represented by a step
function.}
\end{figure*}

The SMBH mass is observed to correlate tightly with both $\sigma_{\rm sph}$
(Tremaine et al.~2002) and $M_{\rm sph}$ (Haering \& Rix~2004), implying
that SMBH growth is regulated by properties of the spheroid. It is
therefore natural to ask {\it which parameter among these two is
responsible for setting the SMBH mass?} We address this question
empirically without specifying a mechanism for the SMBH growth.

Let us suppose first that the $M_{\rm bh}$--$M_{\rm sph}$ relation is the
fundamental one. There would still be a relation between $M_{\rm bh}$ and
$\sigma_{\rm sph}$ since $M_{\rm sph}$ and $\sigma_{\rm sph}$ are related.
Bernardi et al.~(2003) have compiled values of $R_{\rm sph}$, and
$\sigma_{\rm sph}$ for the sample of early-type galaxies in the SDSS. We
can generate a sub-sample of galaxies within a narrow range of $\sigma_{\rm
sph}$ and find the dynamical mass $M_{\rm sph}=V_{\rm sph}^2R_{\rm
sph}/G=2\sigma_{\rm sph}^2R_{\rm sph}/G$ for each, where $V_{\rm sph}$ is
the circular velocity at $R_{\rm sph}$ and we have assumed an isotropic
velocity dispersion $\sigma_{\rm sph}=V_{\rm sph}/\sqrt{2}$. We can then
use the observed relation $M_{\rm bh}\propto M_{\rm sph}$ with its
intrinsic scatter of $\delta_{\rm sph}=0.41\pm0.07$, to find the corresponding
scatter in the $M_{\rm bh}$--$\sigma_{\rm sph}$ relation. The resulting
distributions of residuals in the SMBH mass are plotted on the left panel
of Figure~\ref{fig1}, together with the 1-sigma range in the observed
scatter (grey region) of the $M_{\rm bh}$--$\sigma_{\rm sph}$ relation. If
the $M_{\rm bh}$--$M_{\rm sph}$ relation were fundamental, then the scatter
in the $M_{\rm bh}$--$\sigma_{\rm sph}$ relation would have been
$\delta=0.46$dex, well in excess of the observed value
$\delta=0.275\pm0.05$. This implies that the $M_{\rm bh}$--$M_{\rm sph}$
relation is not fundamental.

On the other hand if we suppose that it is the $M_{\rm bh}$--$\sigma_{\rm
sph}$ relation that is fundamental, then there would still be a relation
between $M_{\rm bh}$ and $M_{\rm sph}$. We generate a sub-sample of
galaxies from Bernardi et al.~(2003) with $M_{\rm sph}=2\sigma_{\rm
sph}^2R_{\rm sph}/G$ in a narrow range near $M_{\rm
sph}=10^{11}M_\odot$. We can then use the observed $M_{\rm
bh}$--$\sigma_{\rm sph}$ relation (with intrinsic scatter
$\delta=0.275\pm0.05$) to find the corresponding scatter in the $M_{\rm
bh}$--$M_{\rm sph}$ relation. The resulting distributions of residuals in
SMBH mass are plotted in the right hand panel of Figure~\ref{fig1},
together with the 1-sigma range in the observed scatter (grey region) of
the $M_{\rm bh}$--$M_{\rm sph}$ relation. If the $M_{\rm bh}$--$\sigma_{\rm
sph}$ relation is fundamental, then the scatter in the $M_{\rm
bh}$--$M_{\rm sph}$ relation should be $\delta_{\rm sph}=0.40$dex, in good
agreement with the observed value $\delta_{\rm sph}=0.41\pm0.07$. This
implies that the $M_{\rm bh}$--$\sigma_{\rm sph}$ relation is more
fundamental than the $M_{\rm bh}$--$M_{\rm sph}$ relation, with the latter
being an incidental consequence of the correlation between $\sigma_{\rm
sph}$ and $M_{\rm sph}$.

\section{Intrinsic Scatter in the $M_{\rm bh}$--$\sigma_{\rm sph}$ Relation and Models for SMBH Evolution}

\begin{figure*}[t]
\epsscale{.8}  \plotone{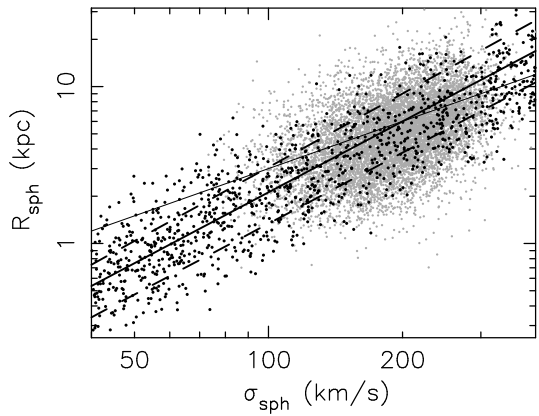}
\caption{\label{fig2} Scatter plot of $R_{\rm sph}$ vs $\sigma_{\rm sph}$
from cooled profiles inside dark matter halos (black-dots). For comparison,
the SDSS sample (Bernardi et al.~2003) is shown by the grey dots. The thick
solid and dashed lines show the mean observed relation $R_{\rm
sph}=6(\sigma_{\rm sph}/200)^{1.5}$ and the level of observed scatter
($\pm0.2$dex). The thin solid line shows a linear relation. }
\end{figure*}

In the previous section we showed that the observed scatter in the $M_{\rm
bh}$--$\sigma_{\rm sph}$ and $M_{\rm bh}$--$M_{\rm sph}$ relations, when
combined with the scatter among spheroid properties implies that the
$M_{\rm bh}$--$\sigma_{\rm sph}$ relation is fundamental for SMBH growth,
with the $M_{\rm bh}$--$M_{\rm sph}$ relation being incidental due to
correlations between spheroid parameters.  This in turn implies that if we
have a model for the properties of the host spheroid, then we can deduce
the mode of SMBH growth by comparing the calculated scatter in the modeled
$M_{\rm bh}$--$\sigma_{\rm sph}$ and $M_{\rm bh}$--$M_{\rm sph}$ relations
with observations.  In this section we constrain the astrophysics of SMBH
growth by computing minimum values for the scatter in the $M_{\rm
bh}$--$\sigma_{\rm sph}$ relation using various models for the regulation
of SMBH growth.

\subsection{Rotation curves, Adiabatic cooling and the Fundamental Plane}
\label{fpsec}

Our goal is to model the scatter in the $M_{\rm bh}$--$\sigma_{\rm sph}$
and $M_{\rm bh}$--$M_{\rm sph}$ relations. To accomplish this goal we must
have a representative model of the host spheroids as well as the scatter in
their parameters. This is achieved by computing the rotation curve that
results from cooling of baryons inside a dark-matter halo.

The virial radius $R_{\rm vir}$ and velocity $V_{\rm vir}$ for a halo
of mass $M_{\rm halo}$ at redshift $z$ are
\begin{equation}
\label{Rvir}
R_{\rm vir} = 109\left(\frac{M_{\rm
halo}}{10^{12}M_\odot}\right)^{1/3}\left(\frac{\Omega_m}{\Omega_m^z}\frac{\Delta_c}{18\pi^2}\right)^{-1/3}\left(\frac{1+z}{2}\right)^{-1}~{\rm
kpc},
\end{equation}
and 
\begin{equation}
V_{\rm vir} = 200\left(\frac{M_{\rm
halo}}{10^{12}M_\odot}\right)^{1/3}\left(\frac{\Omega_m}{\Omega_m^z}\frac{\Delta_c}{18\pi^2}\right)^{1/6}\left(\frac{1+z}{2}\right)^{1/2}~{\rm
km~{s}^{-1}},
\end{equation}
where $\Omega_m^z\equiv[1+(\Omega_\Lambda/\Omega_m)(1+z)^{-3}]^{-1}$,
$\Delta_c=18\pi^2+82d-39d^2$ and $d=\Omega_m^z-1$ (see Barkana \& Loeb~2001
for more details). The relation of the circular velocity at the virial
radius to the velocity dispersion at smaller galactic radii requires
specification of the mass density profile. In this work we adopt the
Navarro, Frenk \& White~(1997, hereafter NFW) profile for the dark matter.
In addition to $V_{\rm vir}$ and $M_{\rm vir}$ the NFW profile is defined
by the concentration parameter $c$, which is the ratio between the virial
radius and a characteristic break radius, $c\equiv r_{\rm vir}/r_s$. The
median value of $c$ depends on halo mass $M_{\rm halo}$ and redshift
$z$. Based on N-body simulations, Bullock et al.~(2001) (see also Wechsler
et al. 2002) have found a median relation for $c$,
\begin{equation}
\label{concparam}
c=7.3\left(\frac{M_{\rm halo}}{10^{12}M_\odot}\right)^{-0.13}\left({1+z\over
2}\right)^{-1},
\end{equation}
with a scatter of $\Delta \log c=0.14$ dex. 

The parameters describing the spheroid are its characteristic radius
$R_{\rm sph}$ and velocity dispersion $\sigma_{\rm sph}$.  In this paper we
assume that the gas available to cool in the halo makes most of the mass
within the effective radius of the spheroid $M_{\rm sph}$ with a density
distribution described by a Hernquist (1990) profile.  The cooling of
baryons modifies the density profile of a galaxy and hence its rotation
curve. In particular, the velocity dispersion in the luminous portion of
the galaxy is substantially larger than would be expected from an NFW
profile. Gnedin et al.~(2004) have studied the contraction of an NFW halo
due to baryon cooling in a cosmological context. They find that traditional
adiabatic contraction does not provide a good fit. However they introduce
an alternative modified contraction model and provide fitting formulae for
contracted profiles as a function of halo mass, concentration parameter,
final characteristic radius for the baryons, and the cooled baryon
fraction. From this contracted profile we can compute the velocity
dispersion $\sigma_{\rm sph}$ at the characteristic radius $R_{\rm sph}$.
However the cooled profile, and hence the value of $\sigma_{\rm sph}$
obtained from the cooled profile depends on the value adopted for $R_{\rm
sph}$. Defining $m_{\rm d}$ to be the fraction of the total galaxy mass
that makes the spheroid (including the cooled baryons), we can break this
degeneracy by identifying the spheroid mass $m_{\rm d}M_{\rm halo}$ with
the effective virial mass $M_{\rm sph}$ through $2\sigma_{\rm sph}^2R_{\rm
sph}/G= M_{\rm sph}\equiv m_{\rm d}M_{\rm halo}$. With this second relation
we are able to solve uniquely for $R_{\rm sph}$ and $\sigma_{\rm sph}$
within a specified dark matter halo (with parameters $M_{\rm halo}$, $c$,
$z$, and $m_{\rm d}$). We choose a probability distribution that is flat in
the logarithm of $m_{\rm d}$ over the range $0.015\leq m_{\rm
d}\leq0.15$. The upper value corresponds to the case where all baryons in
the halo cool to form the mass of the spheroid [so that $M_{\rm
sph}=(\Omega_{\rm b}/\Omega_{\rm m})M_{\rm halo}$], and the range under
consideration spans values smaller by up to an order of magnitude relative
to this extreme case. Using this formalism we generate a sample of model
spheroids. This sample is compared first to the observed fundamental plane
of early-type galaxies (below), and then used to discuss the scatter in
models for the $M_{\rm bh}$--$\sigma_{\rm sph}$ and $M_{\rm bh}$--$M_{\rm sph}$ relations
(\S~\ref{models}).

Figure~\ref{fig2} shows a scatter plot of $R_{\rm
sph}$ vs $\sigma_{\rm sph}$ for cooled profiles with randomly selected
parameters. The distribution of formation redshift is dictated by the
merger trees of halos and corresponding fate of pre-existing galaxies
within these halos (Eisenstein \& Loeb 1996). It is known empirically
that star formation is rather minimal between $0<z<1$ for
elliptical galaxies (Bernardi et al.~2003). 
Since we find that the relation between $R_{\rm sph}$ and $\sigma_{\rm
sph}$ is not very sensitive to the precise probability distribution of
collapse redshifts, we adopt in this figure a distribution of collapse
redshifts that is flat between $1<z<3$.  The resulting points show a
correlation with significant scatter.  For comparison, the relation between
the velocity dispersion $\sigma_{\rm sph}$ of an early type stellar system
and its effective radius $R_e$ for galaxies from the SDSS (Bernardi et
al.~2003) is shown by the grey dots. A fit to this observed correlation has
the form
\begin{equation}
\label{Resigma}
\log{\left(\frac{R_e}{6\mbox{kpc}}\right)} \approx
1.5\log{\left(\frac{\sigma_{\rm sph}}{200\mbox{km}\,\mbox{s}^{-1}}\right)},
\end{equation}
which is plotted as the thick solid line in Figure~\ref{fig2}. The
scatter about the observed relation, $\pm0.2$dex, is bracketed by the
pair of dashed lines. Also plotted to guide the eye is a linear
relation (thin line). On comparison with the results of Bernardi et
al.~(2003), we see that the formalism reproduces the observed behavior
$R_{\rm sph}\propto\sigma_{\rm sph}^{1.5}$.  Moreover the size of the
predicted scatter about the mean relation is $\sim0.17$dex, in good
agreement with the observed value of $\sim0.2$dex. This result is not
very sensitive to the (unknown) distribution chosen for $m_{\rm
d}$. The model does not contain any free parameters. Nevertheless, in
addition to the scatter an power-law slope of the $R_{\rm
sph}$--$\sigma_{\rm sph}$ relation, our prescription also predicts its
normalization.

For completeness we note that the relation described by
equation~(\ref{Resigma}) depends only on dynamical properties and not on
the details of star-formation.  However elliptical galaxies follow a
3-parameter {\em fundamental plane}, with the scatter around the median
$R_{\rm sph}$--$\sigma_{\rm sph}$ relation parameterized by surface
brightness rather than a 2 parameter relation (Djorgovski \& Davis~1987). The
scatter around the {\em fundamental plane} is substantially smaller than
around the $R_{\rm sph}$--$\sigma_{\rm sph}$ relation shown in
Figure~\ref{fig2}. Bernardi et al~(2003) find that the fundamental plane
has the form
\begin{equation}
\label{fp}
R_{\rm sph}\propto\sigma_{\rm sph}^{1.49\pm0.05}I_{\rm sph}^{-0.75\pm0.01},
\end{equation} where the surface
brightness is $I_{\rm sph}\propto L/R_{\rm sph}^2$, and show that this
requires the mass-to-light $\Gamma$ to have a dependence on $R_{\rm sph}$
of the form $\Gamma\propto R_{\rm sph}^{1/3}$.  A complete model of the
{\em fundamental plane} would need to explain this dependence of the
mass-to-light ratio, in addition to the relation $R_{\rm
sph}\propto\sigma_{\rm sph}^{1.5}$. However we do not expect SMBH growth to
depend on $\Gamma$, and do not attempt to model this (orthogonal)
parameter.

Finally, we note that the non-linearity of the relation between $R_{\rm
sph}\propto\sigma_{\rm sph}^{1.5}$ is surprising since it is derived within
the context of CDM dark-matter halos for which the relation between the
virial radius and velocity is $R_{\rm vir}\propto V_{\rm vir}$ at any given
redshift.  In the following section we discuss the growth of SMBHs inside
host spheroids in light of the observed scatters in the $M_{\rm
bh}$--$\sigma_{\rm sph}$ and $M_{\rm bh}$--$M_{\rm sph}$ relations.

\subsection{Models for SMBH Evolution}
\label{models}

\begin{figure*}[t]
\epsscale{2.}  \plotone{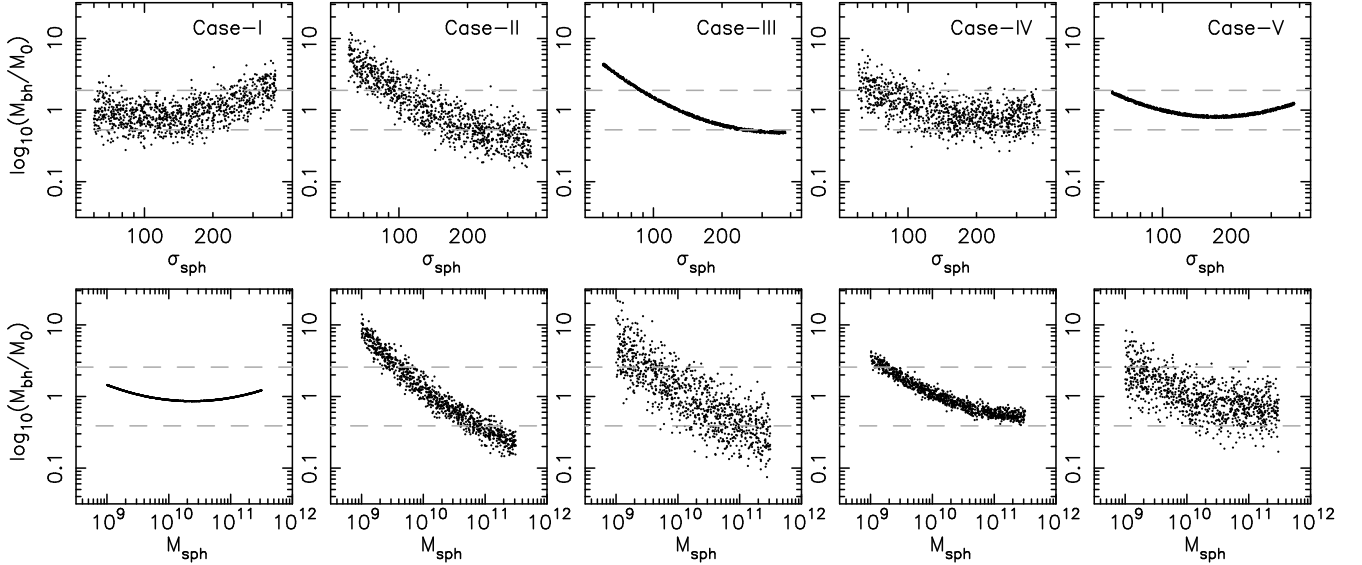}
\caption{\label{fig3} Predicted residual scatter
($\log_{10}(M_{\rm bh}/M_0)$, where $M_0$ is the SMBH mass corresponding to
the mean relation) relative to the best-fit log-quadratic
relations. Results are shown as a function of $\sigma_{\rm sph}$ (upper row) and
$M_{\rm sph}$ (lower row) for models where the SMBH growth is regulated by
properties of the spheroid. The dashed grey lines show the level of observed scatter.}
\end{figure*}

\begin{figure*}[t]
\epsscale{1.8}  \plotone{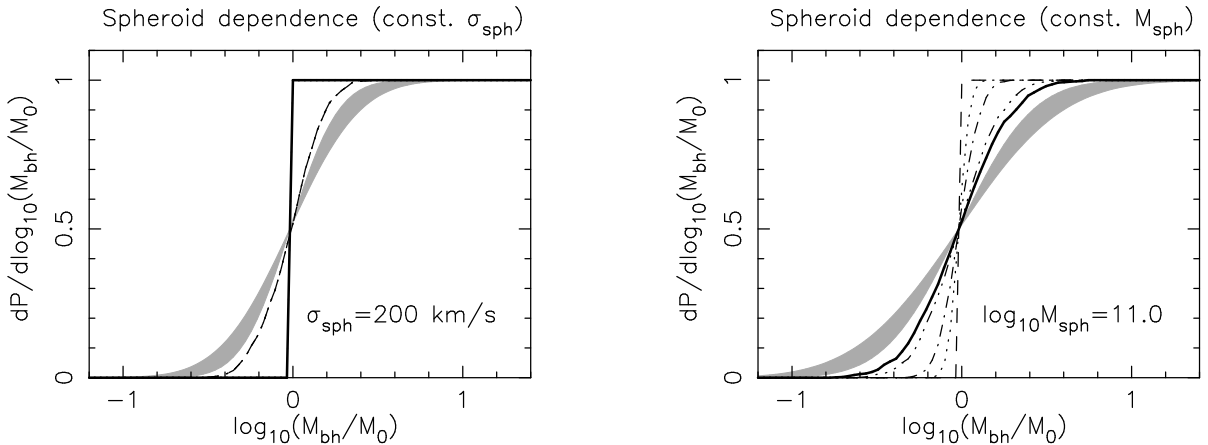}
\caption{\label{fig4} Predicted intrinsic scatter for models where the SMBH
growth is regulated by properties of the spheroid. In each case we
show the predicted cumulative distribution of residuals
($\log_{10}(M_{\rm bh}/M_0)$, where $M_0$ is the SMBH mass
corresponding to the mean relation) at constant $\sigma_{\rm
sph}=200$km/s (left) and constant $M_{\rm sph}=10^{11}M_\odot$
(right). The dashed, dot-dashed, solid, dotted, and dot-dot-dot-dashed
lines correspond to cases I, II, III, IV and V respectively.  In the
left panel the lines for cases-III and V, and for cases-I, II and IV
are coincident. In each panel the range of the observed scatter
(1-sigma) is shown as the shaded grey region for comparison.  }
\end{figure*}

Three observations of the relation between SMBHs and their hosts have
motivated different classes of models to describe the growth and evolution
of SMBHs through accretion. First, the observation of the Magorrian et
al. (1998) relation that the mass of the SMBH follows the mass of the
spheroid has motivated models where the mass of the SMBH accretes a
constant fraction of the available gas following a major merger
(e.g. Haiman \& Loeb 1998; Kauffmann \& Haehnelt~2000). We refer to this
scenario as case-I. Second, the observation of the $M_{\rm
bh}$--$\sigma_{\rm sph}$ relation implies that the SMBH growth is regulated
by the depth of the gravitational potential well of its host spheroid.  We
refer to this scenario as case-II. Third, there is evidence from the
observation of the $M_{\rm bh}$--$\sigma_{\rm sph}$ relation in quasars,
that the $M_{\rm bh}$--$\sigma_{\rm sph}$ relation does not vary with
redshift (Shields et al.~2003). This motivates a revision of case-II, to
include regulation of the SMBH growth over the dynamical time of the system
(e.g. Silk \& Rees~1998; Haehnelt, Natarajan \& Rees~1998; Wyithe \& Loeb
2003).  We refer to this as case-III. On the other hand if momentum rather
than energy is conserved in the transfer of energy in a quasars outflow to
the cold galactic gas, then rather than a SMBH regulated by binding energy,
we have a SMBH mass regulated by the binding energy divided by the virial
velocity (Silk \& Rees 1998; King~2003; Begelman 2004; Murray, Quataert, \&
Thompson 2005). In analogy with cases-II and III, the SMBH mass may be
regulated by the total momentum of the surrounding gas, or by the total
momentum divided by the systems dynamical time. We refer to these as
cases-IV and V, respectively.

We therefore test five hypotheses for each of the $M_{\rm
bh}$--$\sigma_{\rm sph}$ and $M_{\rm bh}$--$M_{\rm sph}$ relations. In the
previous section we described a model that reproduces the observed behavior
of $R_{\rm sph}\propto\sigma_{\rm sph}^{1.5}$, with a scatter of
$\sim0.2$dex. The agreement with the observed projection of the fundamental
plane (Bernardi et al. 2003), gives us confidence that the model provides a
sufficiently accurate framework within which we can discuss the scatter in
the $M_{\rm bh}$--$\sigma_{\rm sph}$ and $M_{\rm bh}$--$M_{\rm sph}$
relations.  Below we list the details of all cases under consideration:

\begin{itemize}

\item Case-I: the mass of the SMBH saturates at a constant fraction of the
mass of the spheroid, $M_{\rm bh}\propto M_{\rm sph}\propto\sigma_{\rm
sph}^2R_{\rm sph}$.

\item Case-II: the mass of the black-hole grows in proportion to the
binding energy of baryons in the spheroid.  Taking a constant fraction
of the cold-gas mass to be material that must be expelled from the
spheroid during the feedback we find that the black-hole mass is
therefore $M_{\rm bh}\propto M_{\rm sph}\sigma_{\rm
sph}^2\propto\sigma_{\rm sph}^4R_{\rm sph}$.

\item Case-III: the mass of the SMBH is determined by the mass for which
accretion at the Eddington limit provides a constant fraction of the
binding energy of the baryons in the spheroid over a constant fraction
of the spheroid's dynamical time.  Thus, the black-hole mass scales as
$M_{\rm bh}\propto E_{\rm b}/(R_{\rm sph}/\sigma_{\rm sph})\propto
M_{\rm sph}\sigma_{\rm sph}^3/R_{\rm sph}\propto\sigma_{\rm sph}^5$.

\item Case-IV: As in case-II, but with the momentum rather than the energy
of the outflow coupling to the gas in the spheroid, yielding $M_{\rm
bh}\propto M_{\rm sph}\sigma_{\rm sph}\propto\sigma_{\rm sph}^3R_{\rm
sph}$.

\item Case-V: As in case-III, but with the momentum rather than the energy
of the outflow coupling to the gas in the spheroid over a dynamical time,
yielding $M_{\rm sph}\sigma_{\rm sph}^2/R_{\rm sph}\propto\sigma_{\rm
sph}^4$.

\end{itemize} 

Studies of the local SMBH inventory suggest that most of mass in SMBHs was
accreted during a luminous quasar phase (e.g. Yu \& Tremaine~2002; Shankar
et al.~2004), with a potentially significant contribution from an
additional dust--obscured accretion phase (Martinez-Sansigre et
al.~2005). If the fraction of obscured quasars is independent of redshift,
then the quasar luminosity function (Fan et al.~2004; Boyle et al.~2000)
can be used as a proxy for the distribution of SMBH formation
redshifts. Using this redshift distribution and the formalism outlined in
the previous sub-section, we can estimate the slope and scatter in the
$M_{\rm bh}$--$\sigma_{\rm sph}$ and $M_{\rm bh}$--$M_{\rm sph}$ relations
under the five different scenarios for the regulation of SMBH growth.

For each of the five cases we perform Monte-Carlo simulations of SMBH
growth. The resulting residuals in the $M_{\rm bh}$--$\sigma_{\rm
sph}$ and $M_{\rm bh}$--$M_{\rm sph}$ relations are plotted in
Figure~\ref{fig3} relative to the best fit log-quadratic relations
(equations~\ref{Msig}-\ref{MM}) as a function of $\sigma_{\rm sph}$
(upper row) and $M_{\rm sph}$ (lower row) respectively. For each of
the five cases, we also compute distributions of residuals (in dex)
relative to the mean SMBH mass at constant $\sigma_{\rm sph}$ and
constant $M_{\rm sph}$.  The resulting distributions are plotted in
Figure~\ref{fig4}. The values of scatter are listed in the 1st and 2nd
columns of Table~\ref{tab1}. The ratio of the scatter at constant
$M_{\rm sph}$ and at constant $\sigma_{\rm sph}$ are listed in
column~3. The power-law slopes $\beta$ and $\beta_{\rm sph}$ in each
case are listed in columns 4 and 5.

\begin{table*}[t]
\begin{center}
\caption{\label{tab1} Predicted scatter and logarithmic slope in the
$M_{\rm bh}$--$\sigma_{\rm sph}$ relation and the $M_{\rm
bh}$--$M_{\rm sph}$ relation.}
\begin{tabular}{ccccccccccc}
\hline 
       & \multicolumn{5}{c}{spheroid} & \multicolumn{5}{c}{halo}\\
       &   $\delta$ & $\delta_{\rm sph}$  & $\delta_{\rm sph}/\delta$ & $\beta$ & $\beta_{\rm sph}$ &   $\delta$ & $\delta_{\rm sph}$  & $\delta_{\rm sph}/\delta$ & $\beta$ & $\beta_{\rm sph}$   \\\hline
Case-I & 0.17 (0.275) & 0.0 (0.22) & 0 (0.8) & 3.4 & 1.0  & 0.24  (0.275) & 0.29 (0.31) & 1.2 (1.1)  & 3.3 & 1.0  \\
Case-II & 0.17 (0.275) & 0.10 (0.24) & 0.6 (0.87) & 5.3 & 1.6 & 0.36 & 0.48 & 1.3  & 5.4 & 1.6  \\
Case-III & {\bf 0.0 (0.275)} & {\bf 0.25 (0.37)} & {\bf $\infty$ (1.35)}  & {\bf 5.0} & {\bf 1.5} & 0.37 & 0.53 &  1.4  & 5.5 & 1.5 \\
Case-IV & 0.17 (0.275) &  0.05 (0.22) & 0.29 (0.8)      & 4.2 & 1.3 & 0.39 & 0.38  & 1.0 & 4.3 & 1.3 \\
Case-V &  {\bf 0.0 (0.275)} &   {\bf 0.20 (0.34)} & {\bf $\infty$ (1.24)}  & {\bf 4.0} & {\bf 1.2} & 0.29 & 0.42   & 1.4  & 4.4 & 1.3  \\
\hline 
\end{tabular}
\end{center}
\end{table*}

From the above discussion we see that under the assumption of a unique
coupling efficiency between the quasar output and the surrounding spheroid,
cases-III and V imply a perfect $M_{\rm bh}$--$\sigma_{\rm sph}$
relation. Cases-I, II and IV do not produce perfect $M_{\rm
bh}$--$\sigma_{\rm sph}$ relations even under this unique coupling
assumption due to the scatter in the $R_{\rm sph}$--$\sigma_{\rm sph}$
relation. Similarly, the scatter in the $R_{\rm sph}$--$\sigma_{\rm sph}$
relation leads to scatter in the $M_{\rm bh}$--$M_{\rm sph}$ relation for
cases-III and V. As a result, cases-III and V predict a scatter in the $M_{\rm
bh}$--$M_{\rm sph}$ relation that is larger than in the $M_{\rm
bh}$--$\sigma_{\rm sph}$ relation, in agreement with observations.

Obviously in these five cases we have neglected many additional sources of
scatter. These include, but are not limited to, the fraction of the
Eddington limit at which the SMBH shines during its luminous phase, the
efficiency of coupling of feedback energy or momentum to the gas in the
host spheroid, and the fraction of the system's characteristic size for
which the dynamical time should be computed. In all five cases the minimum
values of the scatter are smaller than the observed $\delta=0.275$. However
in cases I, II and IV, the models predict a scatter in the $M_{\rm
bh}$--$M_{\rm sph}$ relation that is smaller than in the $M_{\rm
bh}$--$\sigma_{\rm sph}$ relation while the observations show the
opposite. On the other hand if SMBHs are regulated by the binding energy or
momentum of gas in the spheroid per dynamical time of the spheroid
(cases~III \& V), then the minimum scatter in the $M_{\rm
sph}$--$\sigma_{\rm sph}$ relation is reduced to zero. Cases-III and V
therefore predict that the scatter in the $M_{\rm bh}$--$M_{\rm sph}$
relation should be larger than in the $M_{\rm bh}$--$\sigma_{\rm sph}$
relation as observed.

The scatter in the projection of the fundamental plane onto the
$R_{\rm sph}$--$\sigma_{\rm sph}$ plane therefore allows us to differentiate
between SMBH growth that is regulated by the mass (Case-I), binding
energy (Case-II) or momentum of gas (Case-IV) in the spheroid, and
SMBH growth that is regulated by energy or momentum feedback over a
dynamical time of the spheroid (Cases-III and V). In the latter
cases the scatter in the $M_{\rm bh}$--$M_{\rm sph}$ relation is
increased by the large scatter in the spheroid radius, $R_{\rm
sph}$. On the other hand in Cases-III and V, $R_{\rm sph}$ cancels out
in the division of mass by dynamical time in the determination of
$M_{\rm bh}$ at constant $\sigma_{\rm sph}$.

If the predicted value for the minimum scatter in the $M_{\rm
bh}$--$\sigma_{\rm sph}$ relation is smaller than the observed value of
$\delta=0.275$ dex, then there is room in the model for additional random
scatter to account for varying Eddington ratio, outflow geometry, dust
composition, and other factors. In each case we therefore add random
scatter in the formation process at a level which results in the predicted
scatter in the $M_{\rm bh}$--$\sigma_{\rm sph}$ relation being equal to the
observed value of $\delta=0.275$ dex. This value, the corresponding value
predicted for the scatter in the $M_{\rm bh}$--$M_{\rm sph}$ relation
($\delta_{\rm sph}$), and the corresponding ratio ($\delta_{\rm
sph}/\delta$) are listed in parentheses in Table~\ref{tab1}. While cases-I,
II and IV predict ratios of scatter between the $M_{\rm bh}$--$M_{\rm sph}$
and $M_{\rm bh}$--$\sigma_{\rm sph}$ relation that are smaller than unity,
case-III predicts a ratio of $\delta_{\rm sph}/\delta\sim1.4$, and case-V a
ratio of $\delta_{\rm sph}/\delta\sim1.2$, in good agreement with the
observed value ($\delta_{\rm sph}/\delta=1.5\pm0.35$). We therefore
conclude that SMBH growth is likely regulated by feedback over the
spheroid's dynamical time.

A further possible discriminant between models is the power-law slope of
the $M_{\rm bh}$--$\sigma_{\rm sph}$ relations. Cases III \& V, which
satisfy constraints from the observed scatter in the $M_{\rm
bh}$--$\sigma_{\rm sph}$ and $M_{\rm bh}$--$M_{\rm sph}$ relations, produce
power-law $M_{\rm bh}$--$\sigma_{\rm sph}$ relations, with SMBH mass in
proportion to the velocity dispersion raised to the fifth and fourth power
respectively. This could be compared with the power-law value from Tremaine
et al.~(2002) of $\beta=4\pm0.3$ for galaxies with $\sigma_{\rm
sph}\sim200~{\rm km~s^{-1}}$, a comparison which at first sight this
appears to support case-V. However Wyithe~(2005) has found evidence for a
power-law slope that varies with $\sigma_{\rm sph}$ from $\beta\sim4$ near
$\sigma_{\rm sph}\sim200~{\rm km~s^{-1}}$, to $\beta\sim5$ near
$\sigma_{\rm sph}\sim350~{\rm km~s^{-1}}$ (see Eq.~\ref{Msig}).
The slope of the $M_{\rm bh}$--$M_{\rm sph}$ relation is observed to be
close to unity (Haering \& Rix~2004). Of the cases (III and V) that produce
an acceptably small scatter for the $M_{\rm bh}$--$\sigma_{\rm sph}$
relation, we find that case-III yields $\beta_{\rm sph}\sim1.5$, while
case-V leads to a value of $\beta_{\rm sph}\sim1.2$.  However Wyithe~(2005)
also finds evidence for a varying power-law in the $M_{\rm bh}$--$M_{\rm
sph}$ relation, though not at high significance.

Since the models described in this paper each predict power-law relations
between $M_{\rm bh}$ and $\sigma_{\rm sph}$, the residuals with respect to
the log-quadratic fit therefore show curvature as a function of
$\sigma_{\rm sph}$. Cases-III and V both show a power-law slope that agrees
with the best-fit relation at some values of $\sigma_{\rm sph}$, which
renders discrimination between models based on their predicted power-law
slope difficult.  It is beyond the scope of this paper to discuss detailed
models of the $M_{\rm bh}$--$\sigma_{\rm sph}$ relation; however the
observed curvature may point to a turn-over from momentum to energy
conservation in the reaction between the outflow and surrounding
gas. Alternatively, there could be a velocity dependent efficiency of
feedback, which could change the slope of the relation.  In summary, based
on the observed values of $\delta=0.275\pm0.05$, $\delta_{\rm
sph}=0.41\pm0.07$, only models corresponding to Case-III or Case-V are
acceptable. Examples of this class include the models of Silk \&
Rees~(1998); Wyithe \& Loeb (2003); King~(2003), di~Matteo et al.~(2005),
or Murray et al. (2005).

\subsection{Additional scatter from dissipationless mergers after the quasar epoch}

Our model computes the minimum scatter in the $M_{\rm bh}$--$\sigma_{\rm
sph}$ relation during a time when the SMBH can grow via gas accretion. At
late times a SMBH may find itself in an environment where there is no
remaining cold gas. In this regime a merger of two galaxies will proceed
via collisionless dynamics. The two SMBHs may coalesce once they enter the
gravity-wave dominated regime, but since there is no cold gas the SMBH will
not grow during a quasar phase. It would require detailed numerical
simulations to discover whether or not the $M_{\rm bh}$--$\sigma_{\rm sph}$
relation holds following a collisionless merger. However, unlike the
situation where feedback is still at work, in a collisionless merger the
stellar spheroid is not sensitive to the existence of the SMBH (at least at
radii comparable to $R_{\rm sph}$).  Moreover the total SMBH mass is nearly
conserved as long as SMBH binaries coalesce due to interactions with the
surrounding stars and the emission of gravitational radiation (Begelman,
Blandford, \& Rees 1980). The growth of the SMBH and the properties of the
spheroid following a collisionless merger should be independent in the
sense of the $M_{\rm bh}$--$\sigma_{\rm sph}$ relation. Thus, we would
expect collisionless mergers to increase the scatter in the $M_{\rm
bh}$--$\sigma_{\rm sph}$ relation, leaving our estimates of the minimum
scatter intact.

\begin{figure*}[t]
\epsscale{1.5}  \plotone{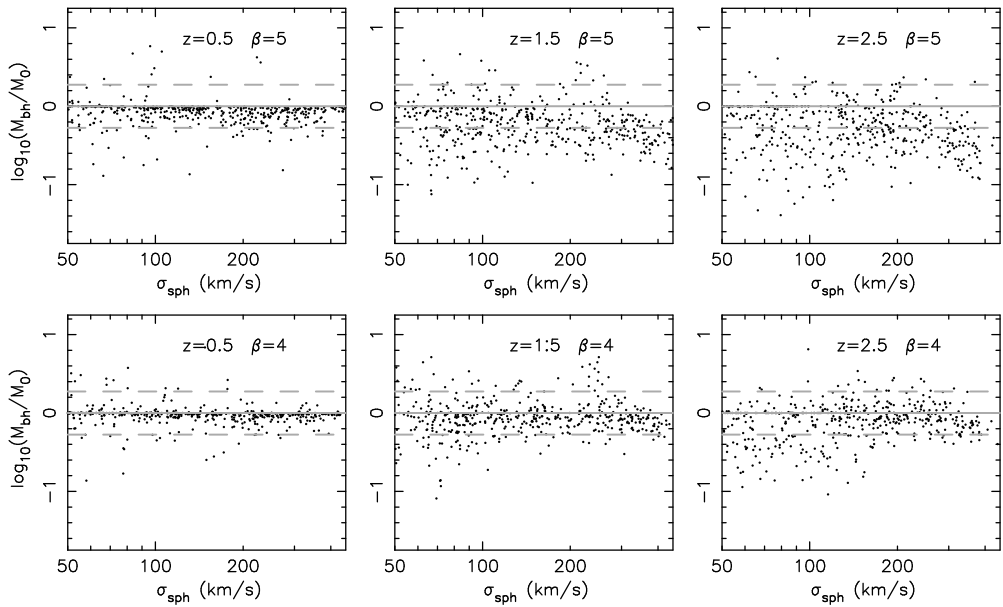}
\caption{\label{fig5} Scatter plots of the predicted residuals
($\log_{10}(M_{\rm bh}/M_0)$, where $M_0$ is the SMBH mass
corresponding to the mean power-law relation at redshift $z$) in $M_{\rm bh}$
that result from collisionless galaxy mergers.  Each point represents
a different realization of a collisionless merger tree beginning with
a perfect power-law $M_{\rm bh}$--$\sigma_{\rm sph}$ correlation at $z=0.5$
(left panel), $z=1.5$ (center panel) and $z=2.5$ (right panel). We
assume Case-III $\beta=5$ (upper row) and Case-V $\beta=4$ (lower
row).}
\end{figure*}

N-body simulations of the behavior of stars as collisionless
particles have been performed by Gao et al.~(2004) who found that the inner mass
density profile (interior to some fixed physical radius) is unaffected by
mergers, implying that the velocity dispersion interior to any given radius
remains the same after a merger.  When two spheroids merge, their combined
stars cover a larger radius (because their mass increases while the inner
mass profile remains unchanged), and this changes the value of $\sigma_{\rm
sph}$.  We may therefore use the fact that the inner profile remains
invariant in order to estimate the scaling between $R_{\rm sph}$ and
$\sigma_{\rm sph}$ in the regime where purely collisionless mergers
occur. Based on the simulations of Gao et al.~(2004), we assume a
universal NFW mass profile for the total mass (dark matter$+$stars)
irrespective of the merger history and find what mergers would do to the
$M_{\rm bh}$--$\sigma_{\rm sph}$ relation when the total mass in stars and
in SMBHs is conserved.  If the inner density profile maintains the NFW
shape of $\rho\propto1/r$, then $\sigma_{\rm sph}^2\propto r\propto M_{\rm
sph}^{1/2}$, i.e. $M_{\rm sph}\propto \sigma_{\rm sph}^4$, similar to the
Faber-Jackson (1976) projection of the fundamental plane of spheroids.

Collisionless mergers will change the average normalization of the $M_{\rm
bh}$--$\sigma_{\rm sph}$ relation. To see why, suppose that we have $N_{\rm
p}$ equal mass progenitors at redshift $z$, each with velocity dispersion
$\sigma_{\rm sph,p}$ and spheroid mass $M_{\rm sph,p}$. At redshift $z$,
the SMBHs obey $M_{\rm bh,p}=C_{\rm p}\sigma_{\rm sph,p}^{\beta_{\rm p}}$,
where $\beta_{\rm p}$ is the slope and $C_{\rm p}$ is a constant. The final
SMBH and spheroid masses at $z=0$ are $M_{\rm bh,f}=N_{\rm p}M_{\rm bh,p}$
and $M_{\rm sph,f}=N_{\rm p}M_{\rm sph,p}$ respectively. Using the above
scaling the final velocity dispersion is $\sigma_{\rm sph,f}=\sigma_{\rm
sph,p}N_{\rm p}^{1/4}$.  We therefore find $M_{\rm bh,f}=N_{\rm p}M_{\rm
bh,p}=N_{\rm p}C_{\rm p}\sigma_{\rm sph,p}^{\beta_{\rm p}}=C_{\rm p}N_{\rm
p}^{1-\beta_{\rm p}/4}\sigma_{\rm sph,f}^{\beta_{\rm p}}$. Thus the
normalization of the $M_{\rm bh}$--$\sigma_{\rm sph}$ relation at fixed
$\sigma_{\rm sph}$ is changed by a factor $\sim N_{\rm p}^{1-\beta_{\rm
p}/4}$ through collisionless mergers.  Note that the slope of the $M_{\rm
bh}$--$\sigma_{\rm sph}$ relation $\beta_{\rm p}$ is preserved if $N_{\rm
p}$ is independent of $\sigma_{\rm sph}$. However the number of progenitors
is a function of halo mass and redshift, and so the change in normalization
could be a function of $\sigma_{\rm sph}$.

In addition to changing the normalization of the $M_{\rm bh}$--$\sigma_{\rm
sph}$ relation, collisionless mergers will also introduce scatter.  In the
limit of a perfect correlation between $M_{\rm sph}$ and $\sigma_{\rm
sph}$, and where the number of progenitors ($N_{\rm p}$) is constant, the
argument in the previous paragraph shows that collisionless mergers would
lead to an $M_{\rm bh}$--$\sigma_{\rm sph}$ relation with no additional
scatter beyond that introduced at the formation redshift. However the
scatter in the $R_{\rm sph}$--$\sigma_{\rm sph}$ correlation combined with
the relation $M_{\rm sph}\propto\sigma_{\rm sph}^2R_{\rm sph}$ implies that
there is $\sim0.2$dex of scatter in $M_{\rm sph}$ at fixed $\sigma_{\rm
sph}$.  Moreover, different galaxies have different merger histories and
therefore a different number of progenitors.  Scatter among the properties
of the initial building blocks at redshift $z$ therefore leads to scatter
in the local $M_{\rm bh}$--$\sigma_{\rm sph}$ relation even if the $M_{\rm
bh}$--$\sigma_{\rm sph}$ relation at $z$ were perfect. This scatter
introduced through collisionless mergers will therefore add scatter to the
local $M_{\rm bh}$--$\sigma_{\rm sph}$ relation beyond that intrinsic to
the formation process itself.

To ascertain the quantitative effect of collisionless mergers on scatter in
the $M_{\rm bh}$--$\sigma_{\rm sph}$ relation, we generate merger trees of
dark-matter halos using the method described in Vollonteri, Haardt \&
Madau~(2003). Based on the merger trees we find the $N_{\rm p}$ progenitor
halos at $z\sim1.5$ that lead to a halo of known mass at $z\sim0$. Using
the formalism outlined in \S~\ref{fpsec} we then determine the values of
$\sigma_{\rm sph}$, $R_{\rm sph}$ and $M_{\rm sph}$ for the spheroids
populating these progenitor halos. We also populate the spheroids with
SMBHs of mass $M_{\rm bh}$ according to the perfect $M_{\rm
bh}$--$\sigma_{\rm sph}$ relations that arise from cases-III and V. The
final SMBH mass residing in the halo at $z=0$ is $M_{\rm
bh,f}=\sum_{i=0}^{N_{\rm p}}M_{{\rm bh},i}$. It is embedded in a spheroid
of mass $M_{\rm sph,f}=\sum_{i=0}^{N_{\rm p}}M_{{\rm sph},i}$. Based on the
above scaling for the inner $\rho\propto1/r$ density profile of an NFW halo
we may estimate the value of velocity dispersion corresponding to the final
spheroid, $\sigma_{\rm sph,f}=\sigma_{\rm sph,0}(M_{\rm sph,f}/M_{\rm
sph,0})^{1/4}$, where $M_{\rm sph,0}$ and $\sigma_{\rm sph,0}$ are the mass
and velocity dispersion of the largest progenitor. In Figure~\ref{fig5} we
show the scatter introduced into a perfect $M_{\rm bh}$--$\sigma_{\rm sph}$
relation originating at $z=0.5$ (left), $z=1.5$ (center) and $z=2.5$
(right) by collisionless mergers between those redshifts and $z=0$. As
discussed above this scatter arises as a result of scatter in the relation
between $\sigma_{\rm sph}$ and $M_{\rm sph}$ in the progenitors. The upper
panels show results for Case-III ($\beta=5$), while the lower panels show
results for Case-V ($\beta=4$). The scatter introduced is roughly
independent of $\sigma_{\rm sph}$ and takes values of $\delta\sim0.1$dex,
$\delta\sim0.2$dex and $\delta\sim0.3$dex for mergers originating at
$z=0.5$, $z=1.5$ and $z=2.5$, respectively.  Thus galaxies that become
devoid of gas at higher redshift lead to a larger scatter in the $M_{\rm
bh}$--$\sigma_{\rm sph}$ relation, because these galaxies undergo more
collisionless mergers by $z=0$ than a galaxy which becomes devoid of gas
only at late times.

The stars that populate massive galaxies appear to be older than those in
low mass galaxies (Kauffmann et al. 2003). The cold gas reservoir that made
these stars must have been depleted at a higher redshift for the
progenitors of high-$\sigma_{\rm sph}$ galaxies.  We might therefore expect
more scatter in the $M_{\rm bh}$--$\sigma_{\rm sph}$ relation at large
$\sigma_{\rm sph}$. For equal mass mergers we have shown that the
normalization of the $M_{\rm bh}$--$\sigma_{\rm sph}$ relation at a fixed
$\sigma_{\rm sph}$ changes by a factor $\sim N_{\rm p}^{1-\beta_{\rm
p}/4}$. Thus, for $\beta_{\rm p}=5$ (Case-III), we find that the amplitude
of the $M_{\rm bh}$--$\sigma_{\rm sph}$ relation should be reduced by collisionless
mergers, while for $\beta_{\rm p}=4$ (Case-V) the amplitude should be
preserved.  This behavior is seen in Figure~\ref{fig5}. Moreover, more
massive galaxies undergo a larger rate of major mergers. Figure~\ref{fig5}
shows that in Case-V the change in normalization of the $M_{\rm
bh}$--$\sigma_{\rm sph}$ relation through collisionless mergers is more
significant in high mass than in low mass galaxies, as
expected. Collisionless mergers could therefore lead to a reduction in the
steepness of the observed $M_{\rm bh}$--$\sigma_{\rm sph}$ relation if
$\beta_{\rm p}>4$.

In summary, in order to satisfy the constraint that the $M_{\rm
bh}$--$\sigma_{\rm sph}$ relation have a scatter at $z=0$ that is smaller
than $\sim0.3$dex, the intrinsic scatter in the relation at the formation
redshift should be smaller than $\sim0.2$ dex, so as to allow for the
additional scatter introduced through collisionless mergers. Cases-III and
V meet this requirement.

\subsection{Redshift dependence of the $M_{\rm bh}$--$\sigma_{\rm sph}$ and $M_{\rm bh}$--$M_{\rm sph}$ relations}

\begin{figure*}[t]
\epsscale{1.8}  \plotone{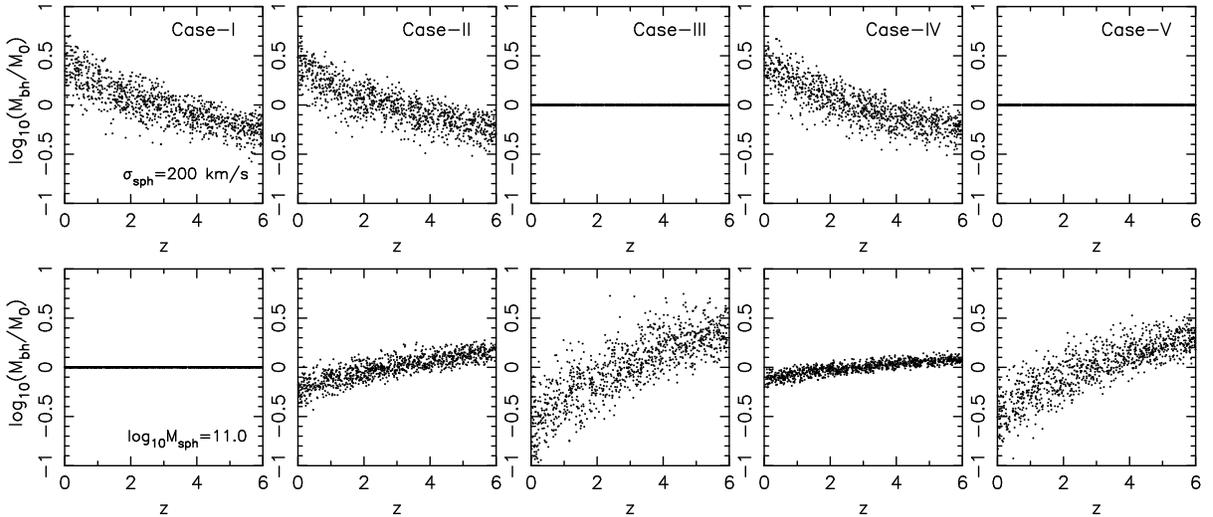}
\caption{\label{fig6} Scatter plots of the predicted residuals
($\log_{10}(M_{\rm bh}/M_0)$, where $M_0$ is the SMBH mass
corresponding to the mean relation) in $M_{\rm bh}$ as a function of
$z$. Results are shown at constant $\sigma_{\rm sph}=200~{\rm
km~s^{-1}}$ (upper row) and constant $M_{\rm sph}=10^{11}M_\odot$ (lower row)
for models where SMBH formation is governed by spheroid properties. In
each of the five cases shown the residuals are relative to the mean
relation at $z=3$.}
\end{figure*}

Recent evidence suggests that the $M_{\rm bh}$--$\sigma_{\rm sph}$
relation is preserved out to high redshift (Shields et al.~2003),
while SMBHs make a larger fraction of their host spheroid mass at
higher redshift (Rix et al.~2001; Croom et al.~2004; Walter et al.~2004).
This behavior is reproduced in models with scenarios of the form
case-III or case-V. Figure~\ref{fig6} shows a scatter plot of the
predicted residuals in $M_{\rm bh}$ vs $z$ at constant $\sigma_{\rm
sph}=200~{\rm km~s^{-1}}$ (upper row) and $M_{\rm sph}=10^{11}M_\odot$
(lower row) for each of the five models. The residuals are normalized
relative to the mean relation at $z=3$ in each case. While there is no
evolution in the $M_{\rm bh}$--$\sigma_{\rm sph}$ relation for
cases-III and V, we see that SMBHs are predicted to be an order of
magnitude more massive with respect to their host spheroid at $z\sim6$
than they are at $z\sim1$ in agreement with observations.  In
contrast, models for case-I, II and IV predict that the SMBH mass
should decrease by an order of magnitude between $z=0$ and $z=6$ at
constant $\sigma_{\rm sph}$, while not evolving significantly at
constant $M_{\rm sph}$. The observed evolution of SMBH mass with
redshift therefore supports cases III and V as the scenario for SMBH
growth. This result supports the findings of \S~\ref{models} based on
the scatter in local relations.

\subsection{Do dark-matter halos play a role in SMBH evolution?}

\begin{figure*}[t]
\epsscale{2.}  \plotone{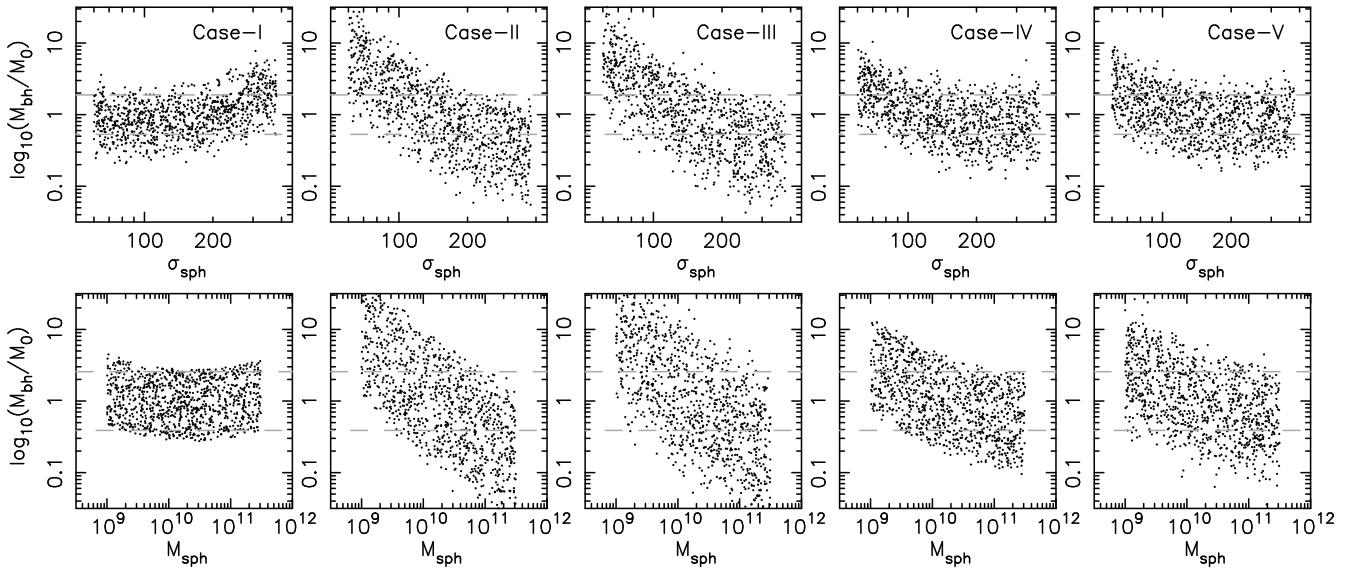}
\caption{\label{fig7} Predicted scatter of residuals ($\log_{10}(M_{\rm bh}/M_0)$ where $M_0$ is
the SMBH mass corresponding to the mean relation) relative to the
best-fit log-quadratic relations. Results are shown as a function of
$\sigma_{\rm sph}$ (upper row) and $M_{\rm sph}$ (lower row) for
models where the SMBH growth is regulated by properties of the dark
matter halo. The dashed grey lines show the level of observed
scatter.}
\end{figure*}

Attempts to reproduce the observed luminosity function of quasars associate
the mass of the SMBH with the properties of the host dark-matter halo. This
paradigm allows the abundance of SMBHs to be traced either in semi-analytic
or numerical models (e.g. Haiman \& Loeb~1998; Haehnelt, Natarajan \&
Rees~1998; Wyithe \& Loeb~2003; Vollonteri, Haardt \& Madau~2002). Indeed,
Ferrarese~(2001) has inferred a relation between the masses of SMBHs and
their host dark-matter halo. {\it Is it possible that the halo rather than
the spheroid regulates SMBH growth?}

Since we have computed spheroid properties within a specified dark matter
halo, we are in a position to discuss the role of the dark matter halo in
regulating the SMBH growth.  In addition to the five cases listed above for
SMBH growth within a spheroid, we also try five analogous cases for the
formation of SMBHs governed by dark matter halo properties.  For each of
the additional five cases, we compute the distribution of residuals (in
dex) relative to the mean relation as a function of $\sigma_{\rm sph}$ and
$M_{\rm sph}$ via a Monte-Carlo algorithm for SMBH formation, and calculate
the variance at each of constant $\sigma_{\rm sph}=200~{\rm km~s^{-1}}$ and
constant $M_{\rm sph}=10^{11}M_\odot$.  Below we list the details of each
case. The resulting distributions of residuals are plotted in
figure~\ref{fig7}. The values of scatter are listed in the 6th and 7th
columns of Table~\ref{tab1}. The ratio of the scatter at constant $M_{\rm
sph}$ and at constant $\sigma_{\rm sph}$ are listed in column~8. Values for
the slopes $\beta$ and $\beta_{\rm sph}$ are listed in columns 9 and 10.

\begin{figure*}[t]
\epsscale{.8}  \plotone{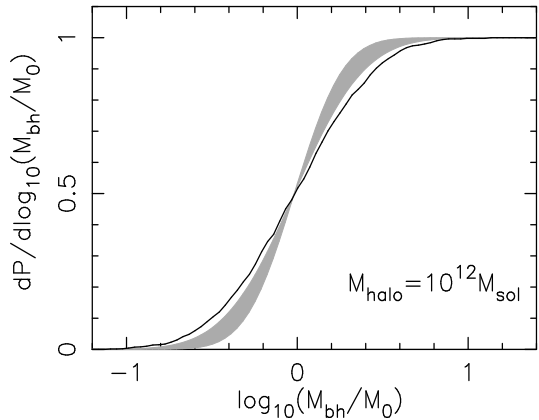}
\caption{\label{fig8} The predicted distribution of intrinsic residuals ($\log_{10}(M_{\rm bh}/M_0)$ where $M_0$ is
the SMBH mass corresponding to the mean relation) in the local $M_{\rm bh}$ at $M_{\rm halo}=10^{12}M_\odot$. The $M_{\rm bh}$--$M_{\rm halo}$ relation has larger intrinsic scatter than the $M_{\rm bh}$--$\sigma_{\rm sph}$ relation.}
\end{figure*}

\begin{itemize}

\item Case-I: the mass of the SMBH forms from a constant fraction of the
baryonic component of the halo mass, $M_{\rm bh}\propto ({\Omega_{\rm
b}}/{\Omega_{\rm m}})M_{\rm halo}$.

\item Case-II: the mass of the black-hole grows in proportion to the
binding energy of baryons in the halo.  For an NFW profile with a
concentration parameter $c$ we get $M_{\rm bh}\propto ({\Omega_{\rm
b}}/{\Omega_{\rm m}})M_{\rm halo} V_{\rm vir}^2 f_{\rm c}$, where
\begin{equation}
\label{fc}
f_{\rm c} = \frac{c}{2}\frac{1-1/(1+c)^2-2\ln(1+c)/(1+c)}{[c/(1+c)-\ln(1+c)]^2}.
\end{equation}

\item Case-III: the mass of the SMBH is determined by the mass for which
accretion at the Eddington limit provides the binding energy of baryons in
the halo over a constant fraction of the halo's dynamical time (Wyithe \&
Loeb~2003). We therefore have $M_{\rm bh}\propto \frac{(\Omega_{\rm
b}/\Omega_{\rm m}) M_{\rm halo} V_{\rm vir}^2 f_{\rm c}}{R_{\rm vir}/V_{\rm
vir}}\propto f_{\rm c} V_{\rm vir}^5$.

\item Case-IV: same as case-II, but with the momentum rather than the
energy of the outflow coupling to the gas in the spheroid. We then find
$M_{\rm bh}\propto ({\Omega_{\rm b}}/{\Omega_{\rm m}})M_{\rm halo} V_{\rm vir}
f_{\rm c}$

\item Case-V: same as case-IV, but with the momentum rather than the energy
of the outflow coupling to the gas in the spheroid over a dynamical
time. We then find $M_{\rm bh}\propto \frac{(\Omega_{\rm
b}/\Omega_{\rm m}) M_{\rm halo} V_{\rm vir} f_{\rm c}}{R_{\rm vir}/V_{\rm
vir}}\propto f_{\rm c} V_{\rm vir}^4$.

\end{itemize}

In these five cases we have again neglected many possible causes of
scatter. However in case-I, where the SMBH growth is regulated by halo
properties, the minimum value of the scatter in the $M_{\rm
bh}$--$\sigma_{\rm sph}$ relation is slightly smaller than the
observed $\delta=0.275$. We have added random scatter to the model for
case-I in order to bring the predicted scatter in the $M_{\rm
bh}$--$\sigma_{\rm sph}$ up to the observed value. This value and the
corresponding prediction for $\delta_{\rm sph}$ and
$\delta/\delta_{\rm sph}$ are listed in parentheses. Case-I predicts a
scatter in the $M_{\rm bh}$--$M_{\rm sph}$ relation that is similar to
the $M_{\rm bh}$--$\sigma_{\rm sph}$ relation while the observations
suggest the latter to be significantly smaller.  While case-I cannot
be ruled out based only on the predicted minimum scatter in the
$M_{\rm bh}$--$\sigma_{\rm sph}$ relation at the formation redshift,
the small allowance for additional expected scatter from the
aforementioned astrophysical sources renders it unlikely, particularly
when the additional scatter of $0.1-0.3$dex from collisionless mergers
at low redshift is accounted for.  
In cases II, III, IV and V where the SMBH growth is regulated by halo
properties, the minimum values of the scatter in the $M_{\rm
bh}$--$\sigma_{\rm sph}$ relation are larger than the observed
$\delta=0.275$. 

While scatter within models of spheroid regulated SMBH
growth is not sensitive to the distribution of $m_{\rm d}$, the predicted
scatter in models of halo regulated SMBH growth decreases as the assumed
range of $m_{\rm d}$ decreases. However, in order to reduce the predicted
scatter below the $\sim0.2$dex threshold at the formation redshift (which
would allow for additional scatter to be introduced through collisionless
mergers), we find that the allowed range around $m_{\rm d}=0.05$ would
need to be smaller than $\pm0.025$, which is implausibly narrow. We
therefore conclude that it is the spheroid rather than the dark-matter
halo which drives the evolution of SMBH mass.

\subsection{The $M_{\rm bh}$--$M_{\rm halo}$ relation}

We have demonstrated that the tight relation between $M_{\rm bh}$ and
$\sigma_{\rm sph}$ implies that it is the spheroid rather than the halo
which governs the growth of SMBHs. However it is clear that since there is
an $M_{\rm bh}$--$\sigma_{\rm sph}$ relation, and since larger halos will,
on average, host bulges with larger central velocity dispersions, there
should also be a correlation between SMBH and halo mass. Ferrarese~(2001)
has found such a relation. Since it is not possible to measure the dark
matter halo mass directly, halo masses for galaxies in the local sample
were inferred via a maximum circular velocity estimated from $\sigma_{\rm
sph}$ based on an empirical relation. It is therefore difficult to estimate
the scatter in the $M_{\rm bh}$--$M_{\rm halo}$ relation
observationally. Here we predict the scatter in the $M_{\rm bh}$--$M_{\rm
halo}$ relation at a fixed value of $M_{\rm halo}$. We compute the
distribution of residuals via a Monte-Carlo method as before. We choose
$M_{\rm halo}=10^{12}M_\odot$ and find the distribution of values for
$\sigma_{\rm sph}$, and hence the distribution of $M_{\rm bh}$ using the
observed $M_{\rm bh}$--$\sigma_{\rm sph}$ relation. The resulting
distribution is plotted in Figure~\ref{fig8}. The variance is $\delta_{\rm
halo}=0.4$ dex. Thus the tightness of the $M_{\rm bh}$--$\sigma_{\rm sph}$
relation suggests that the $M_{\rm bh}$--$M_{\rm halo}$ correlation is
incidental to the fundamental relation between the SMBH and its host
spheroid. Note that this variance is computed at the time of the SMBH
formation. The surrounding dark matter halo could continue to grow after
the supply of cold gas to the SMBH had ceased. This is consistent with our
conclusion that SMBHs grow in proportion to the properties of the spheroid
rather than the halo. Indeed one finds massive dark-matter halos in X-ray
clusters, which must have increased their velocity dispersion well beyond
the corresponding SMBH growth (due to the lack of cooling flows in cluster
cores). This late-time growth of dark-matter halos increases considerably
the scatter in the $M_{\rm bh}$--$M_{\rm halo}$ relation.

\section{conclusion}

We have investigated the implications of intrinsic scatter in the local
relations involving SMBHs for models of SMBH formation. Using the sample of
spheroid properties from SDSS (Bernardi et al.~2003) we first examined
empirically the fundamental parameter describing SMBH growth. The observed
scatter in the $M_{\rm bh}$--$\sigma_{\rm sph}$ relation is
$\delta=0.275\pm0.05$, while the $M_{\rm bh}$--$M_{\rm sph}$ relation has a
larger observed scatter of $\delta_{\rm sph}=0.41\pm0.07$. Assuming that
the $M_{\rm bh}$--$M_{\rm sph}$ relation is fundamental, we use the SDSS
spheroid sample to compute the resulting scatter in the $M_{\rm
bh}$--$\sigma_{\rm sph}$ relation. We find that this procedure results in a
scatter of the $M_{\rm bh}$--$\sigma_{\rm sph}$ relation that is too large
to be reconciled with observation. Alternatively, one might assume that
SMBH growth is determined by $\sigma_{\rm sph}$ rather than $M_{\rm
sph}$. In this case we used the SMBH sample to compute the resulting
scatter in the $M_{\rm bh}$--$M_{\rm sph}$ relation, and found agreement
with the observed scatter. We therefore conclude that SMBH growth is
governed by $\sigma_{\rm sph}$, and that the observed correlation between
$M_{\rm bh}$ and $M_{\rm sph}$ is a by-product of the relation between
$M_{\rm sph}$ and $\sigma_{\rm sph}$.

Theoretical models for SMBH formation must reproduce several observational
constraints: {\it (i)} the scatter in the local $M_{\rm bh}$--$\sigma_{\rm
sph}$ relation is $\delta=0.275\pm0.05$ dex, implying that at the time of
formation the scatter should be smaller than $\sim0.2$dex to allow for
additional scatter introduced by collisionless mergers of galaxies since
$z\sim1$ or earlier; {\it (ii)} the scatter in the $M_{\rm bh}$--$M_{\rm
sph}$ relation is larger than in the $M_{\rm bh}$--$\sigma_{\rm sph}$
relation (this result is maintained as additional scatter from
collisionless mergers is introduced after SMBH formation); and {\it (iii)}
The $M_{\rm bh}$--$\sigma_{\rm sph}$ relation is preserved out to high
redshift. We find that these constraints are only met by models where SMBH
growth is regulated by feedback on the gas feeding the SMBH over the
spheroid dynamical time. Other models lead to scatter in the $M_{\rm
bh}$--$\sigma_{\rm sph}$ relation that are too large or scatter in the
$M_{\rm bh}$--$M_{\rm sph}$ relation that is smaller than the $M_{\rm
bh}$--$\sigma_{\rm sph}$ relation. In addition, other models lead to a SMBH
mass that drops with increasing redshift at a fixed velocity dispersion.
The feedback in successful models can be either in the form of energy or
momentum transfer between the quasar and the galactic gas, leading to
power-law slopes in the $M_{\rm bh}$--$\sigma_{\rm sph}$ relation of
$\beta=4$ or $\beta=5$, respectively. Both of these slopes are permitted
by the local sample (Wyithe~2005).

The above constraints do {\it not} permit SMBH growth to be governed by the
properties of the dark-matter halos. Such models lead to scatter in the
$M_{\rm bh}$--$\sigma_{\rm sph}$ relation that are too large and/or scatter
in the $M_{\rm bh}$--$M_{\rm sph}$ relation that is smaller than the
$M_{\rm bh}$--$\sigma_{\rm sph}$ relation. The relation between $M_{\rm
bh}$ and the halo mass (Ferrarese~2001) has a large scatter ($\sim0.4$dex)
and is most likely a by-product of the correlation between halo mass and
$\sigma_{\rm sph}$.

\acknowledgements 

J.S.B.W. acknowledges the support of the Australian Research Council.  This
work was supported in part by NASA grants NAG 5-13292, NNG-05-GH54G, and by
NSF grants AST-0071019, AST-0204514 (for A.L.).

\end{document}